\documentclass[twocolumn,showpacs,preprintnumbers,amsmath,amssymb]{revtex4}

\usepackage{graphicx}
\usepackage{dcolumn}
\usepackage{bm}

\begin{document}

\title{Coupling between two singing wineglasses}

\author{Tal Arane}
\author{Ana K. R. Musalem}
\author{Moti Fridman}


\affiliation{%
Dept. of Physics of Complex Systems, Weizmann Institute of
Science, Rehovot 76100, \\Israel
}%

\date{\today}

\begin{abstract}
Coupling between two singing wineglasses was obtained and
investigated. Rubbing the rim of one wineglass produce a tone and
due to the coupling induces oscillations on the other wineglasses.
The needed coupling strength between the wineglasses to induce
oscillations as a function of the detuning was investigated.
\end{abstract}

\pacs{43.20+g, 62.30+d}
\maketitle                           


A wineglass can be made to "sing" by gently rubbing its rim with a
moist finger. The friction between the finger and the rim of the
wineglass causes the wineglass to oscillate, to produce a loud,
pure tone \cite{Thomas}. The frequency of the oscillations depends
on the volume of liquid inside the wineglass, whereby adding
liquid increases the fluid pressure that retards the wineglass
vibration so as to lower the frequency
\cite{YihYuh}-\cite{YihYuh2}.

Here we investigate how two singing wineglasses can be coupled to
each other. Specifically, we show that when the natural
oscillating frequency of each wineglass is comparable to each
other, it is relatively easy to induce oscillations of the same
frequency and phase from one wineglass to the other. However, when
their natural oscillation frequencies differ, it is more difficult
to induce the oscillations, if at all. Such coupling plays an
important role in many ensembles such as coupled bubbles
\cite{bubble}, coupled lasers \cite{laser}, etc.

\begin{figure}[h]
\centerline{\includegraphics[width=8cm]{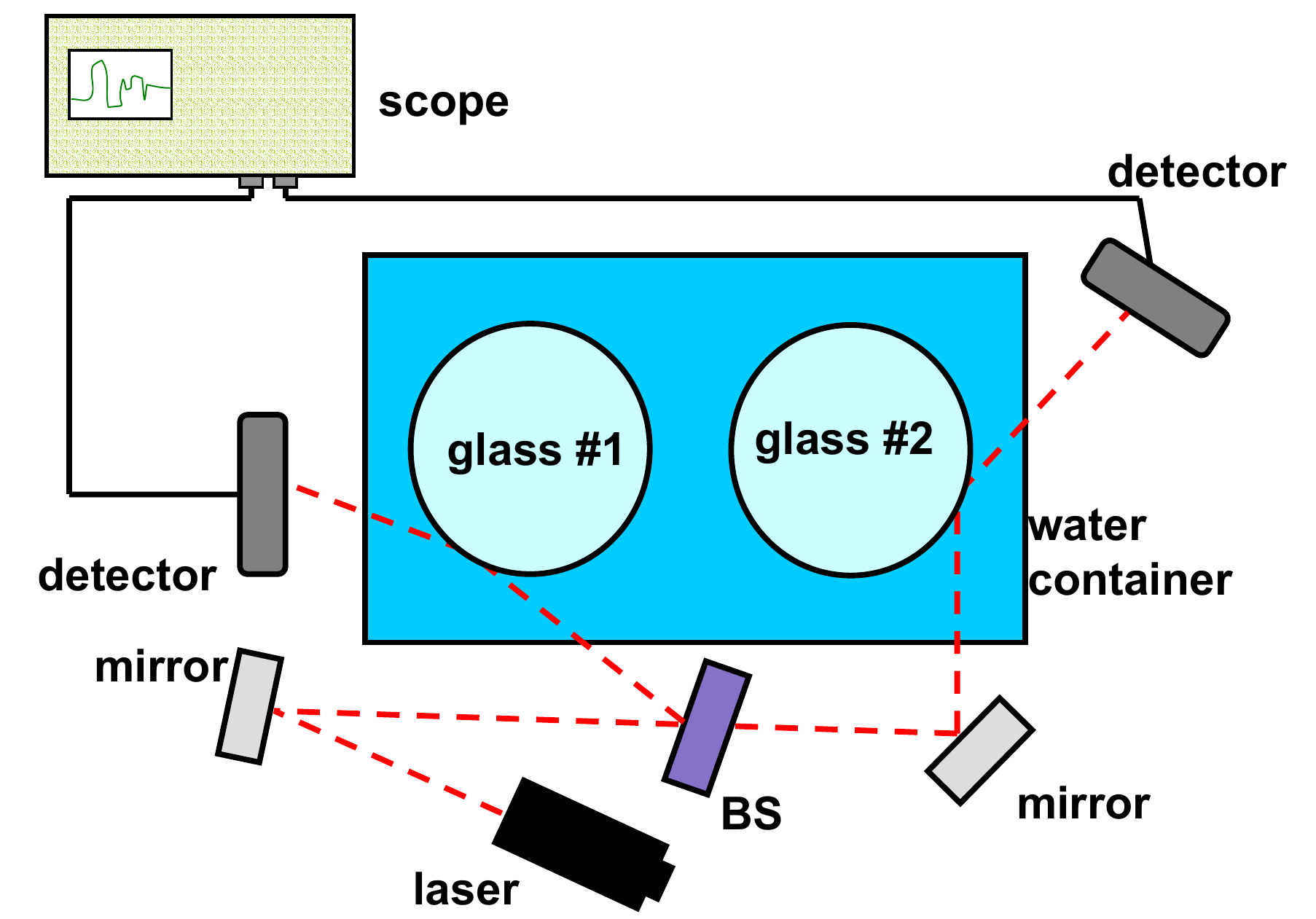}}
\caption{\label{TwoGlasses} Experimental arrangement for
investigating the coupling between two wineglasses.}
\end{figure}

The experimental arrangement is schematically shown in
Fig.~\ref{TwoGlasses}. Two wineglasses were submerged in a water
container, and the oscillating frequency of each wineglass was
obtained by illuminating them with laser beams derived from a HeNe
laser and detecting the reflected light. In this arrangement the
ease of introducing oscillations from one wineglass to the other,
i.e, the coupling strength between the wineglasses depends on the
distance between them, on the water level in the container and on
the difference between the natural oscillation frequencies of the
wineglasses (detuning).

\begin{figure}[h]
\centerline{\includegraphics[width=10cm]{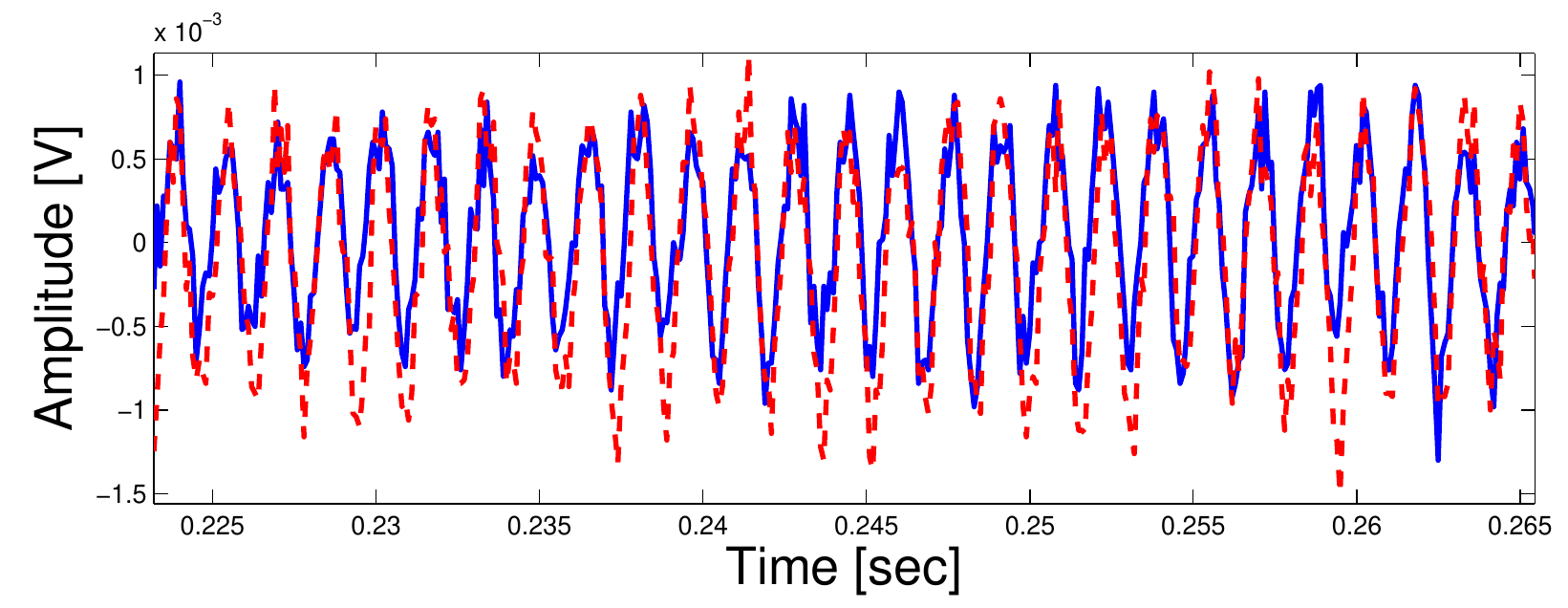}}
\caption{\label{detectors} Oscillation frequencies and amplitudes
for two coupled wineglasses. Dashed curves denote the driving
wineglass and solid curves the driven. The level of the water in
the container was 13cm. The distance between the wineglasses was
5mm.}
\end{figure}

We began our experiments by measuring the natural frequency of
each wineglass as a function of volume of water inside them. We
found, as expected, that the natural oscillating frequency
decreases from 700HZ to 450HZ as the volume of water inside them
increases. Then, we investigated the behavior and content of the
induced oscillations as compared to the detuning oscillations,
when the water level inside each was the same as the other. This
was done by rubbing the rim of only one wineglass to produce a
natural oscillation frequency with a certain amplitude and
measuring the amplitude of the oscillation frequency that is
induced in the other wineglass. The coupling strength between the
wineglasses corresponds to the ratio of amplitudes of the
oscillation frequencies. A representative example of the
oscillation frequencies, amplitudes and phases for the rubbed
wineglass (driving wineglass) and the unrubbed wineglass (driven
wineglass) is shown in Fig~\ref{detectors}. As evident, the
frequencies and phase of both the driving and induced oscillations
are essentially identical. Clearly indicating that frequency
locking and phase locking accrue.

\begin{figure}[th]
\centerline{\includegraphics[width=8cm]{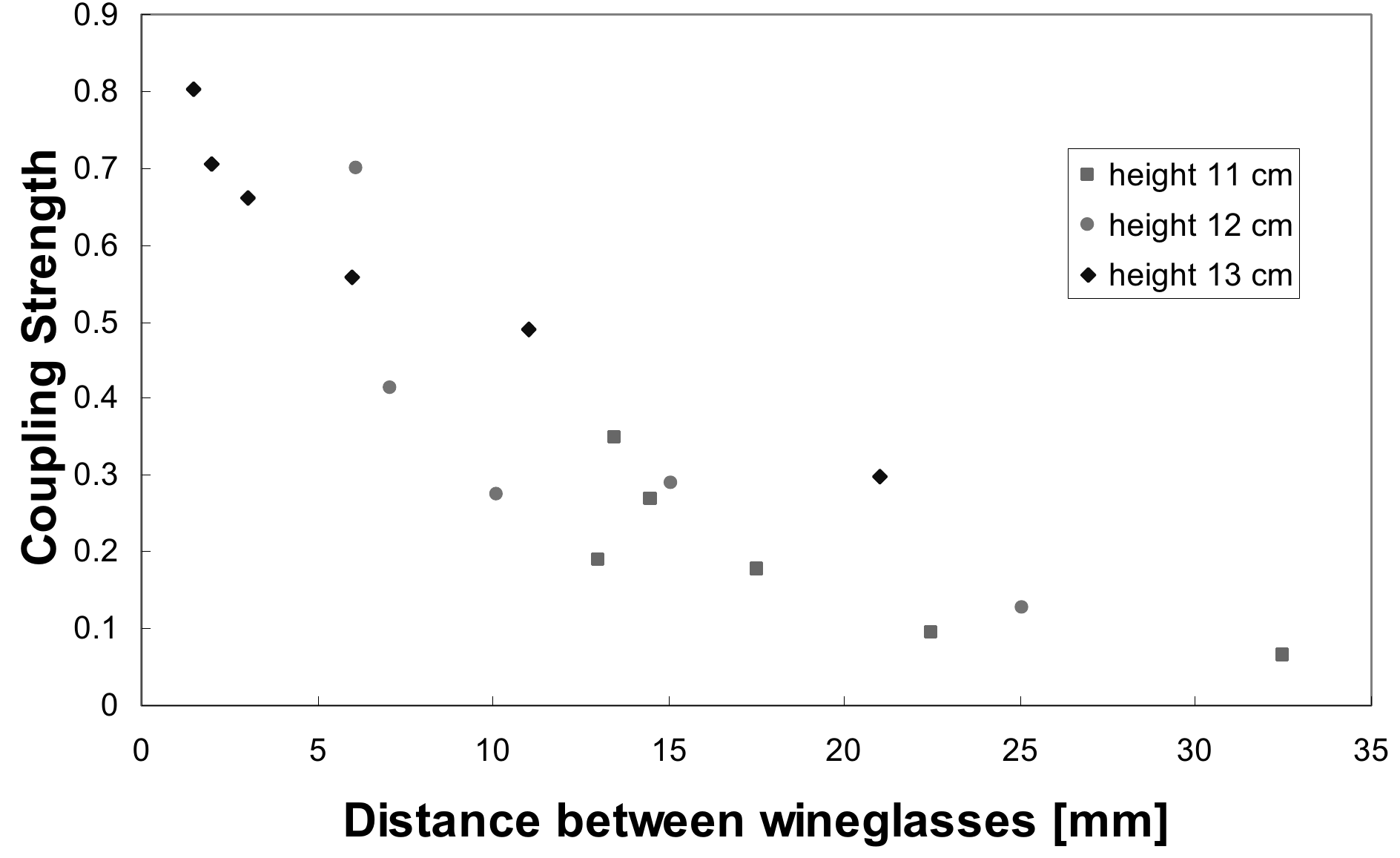}}
\caption{\label{CouplingStrength} Coupling strength between two
wineglasses as a function of the distance between them.}
\end{figure}

\begin{figure}[h]
\centerline{\includegraphics[width=8cm]{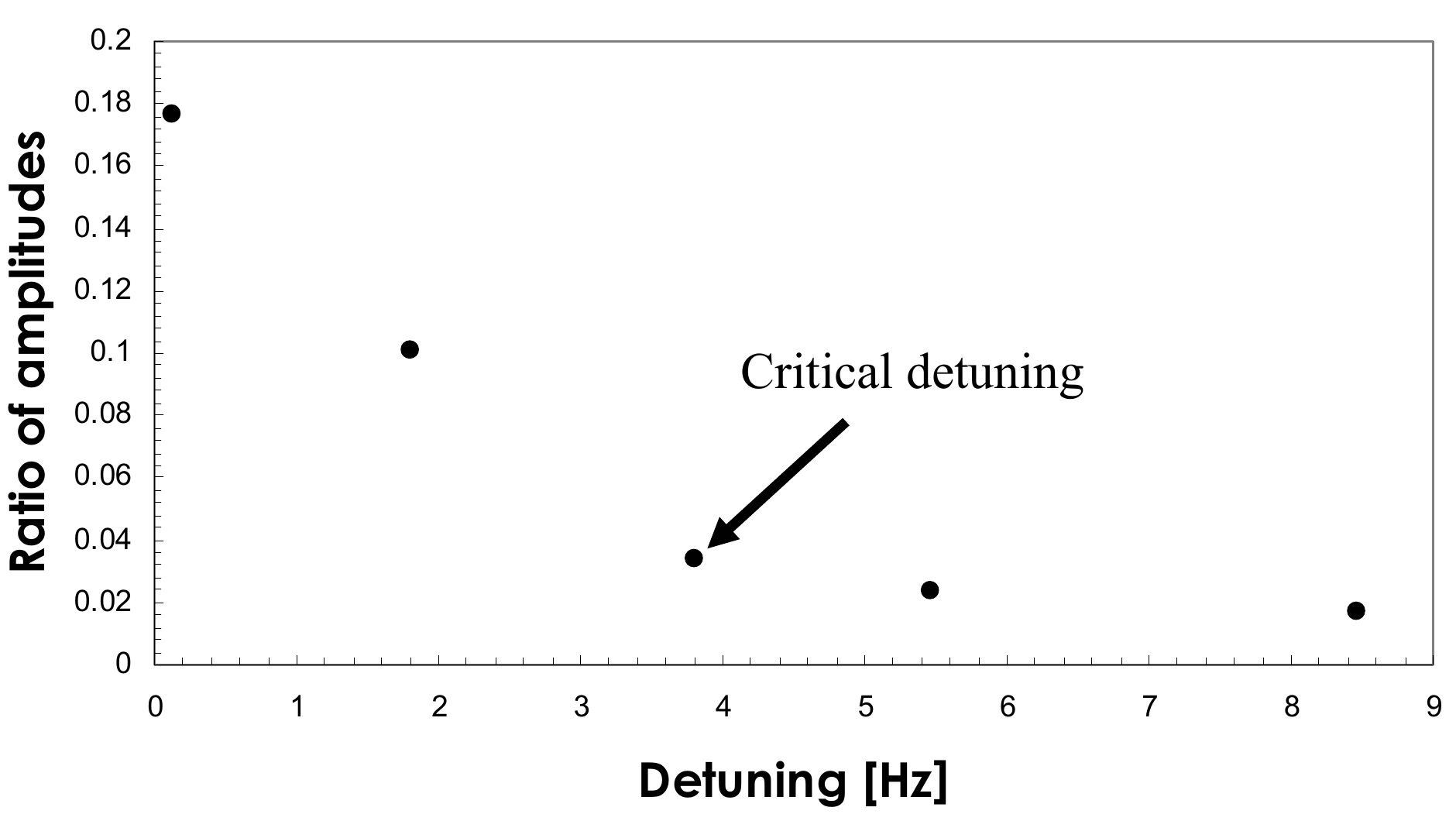}}
\caption{\label{CriticalPoint} Ratio of amplitudes of the two
wineglasses' oscillations as a function of detuning between them.
The distance between the wineglasses was $25 mm$, and the water
level in the container was $12 cm$.}
\end{figure}

\begin{figure}[h]
\centerline{\includegraphics[width=8cm]{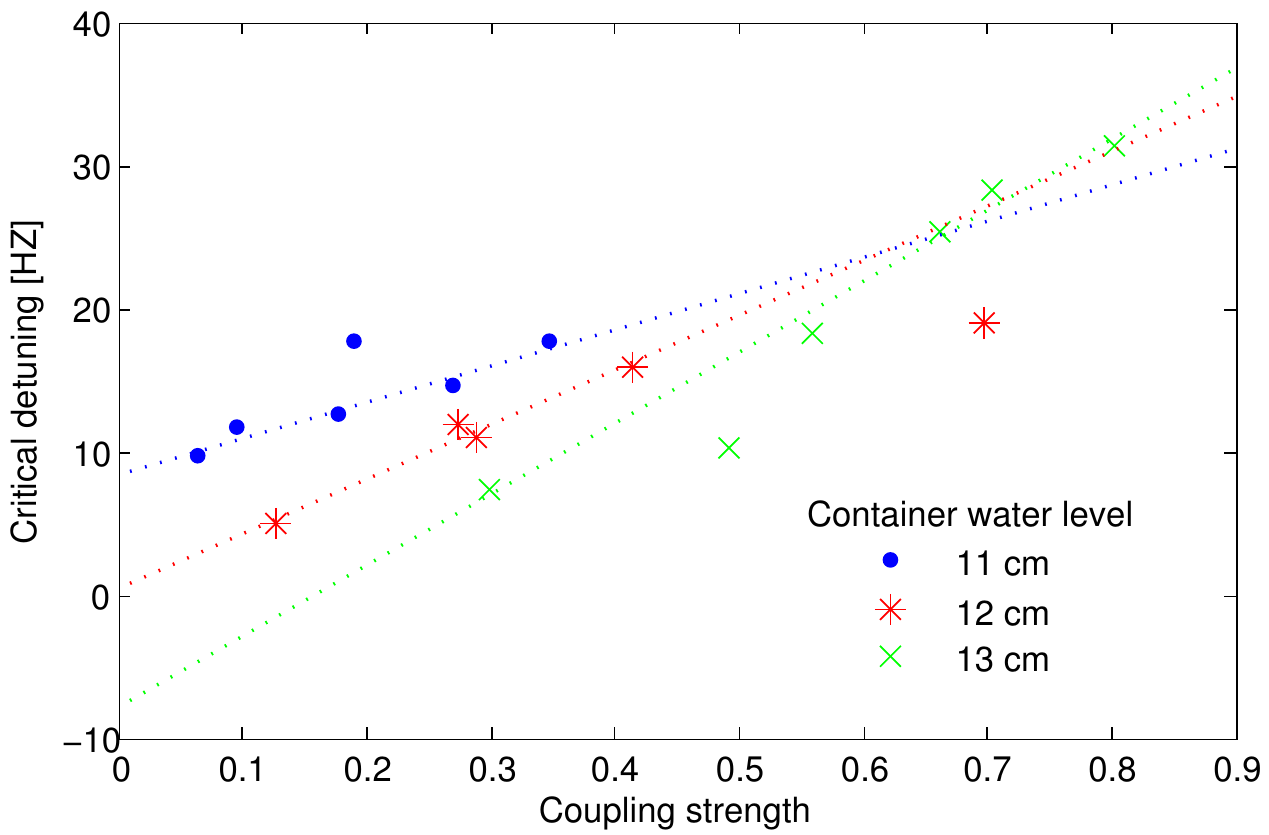}}
\caption{\label{CriticleCoupling} Critical detuning as a function
of coupling strength for different water levels in the container.}
\end{figure}

In order to determine the influence of distance between
wineglasses and detuning, we repeated the measurement above at
different distances and different detuning conditions. The results
are presented in
Figs~\ref{CouplingStrength}-~\ref{CriticleCoupling}. Figure
\ref{CouplingStrength} shows the coupling strength as a function
of distance between the wineglasses, for different levels of water
in the container. As evident, the coupling strength decreases
significantly as the distance between the wineglasses increases.
On the other hand, the influence of water levels of water in the
container is not significant.

Figure 4 shows the coupling as a function of the detuning between
them, for a certain level of water in the container (13cm) and a
certain distance between the wineglasses (25mm). The detuning was
achieved by simply varying the volume of water inside one of the
wineglasses. As evident, the induced frequency amplitude decreases
as the detuning is increased. There is a point of critical
detuning which indicates that induced oscillations can no longer
be achieved in the driven wineglasses at a certain coupling
strength.

Finally, Fig.~\ref{CriticleCoupling} shows the critical detuning
as a function of coupling strength, for different levels of water
in the container. As evident, when the coupiling strength which
induce oscillations in the driven wineglass increases, the
critical detuning increases as well. Also, when the level of water
in the contained increases the slops of the curves increase,
probably due to changes in the damping strength which depend on
the water level in the container

To conclude, the critical detuning of two wineglasses is directly
related to the coupling strength between them and vice versa. For
higher coupling strength the critical detuning is higher. For
large detuning, it is necessary to increase the coupling strength
in order to ensure that induced oscillations will accure in the
driven wineglass.

\end{document}